\documentclass[useAMS,usenatbib]{mn2e}
\usepackage{graphicx,epsfig}
\title[Kinetic effects in solar corona ...]{Missing bits of the solar jigsaw puzzle: 
small-scale, kinetic effects in coronal studies}
\author[David Tsiklauri]{David Tsiklauri\\
Joule Physics Laboratory, Newton Building, University of Salford, Greater Manchester, M5 4WT, UK}
\begin{document}
\date{Accepted xxxx December xxx. Received xxxx December xx; in original form xxxx October xx}
\pubyear{2009}
\maketitle
\label{firstpage}

\begin{abstract}
The solar corona, anomalously hot outer atmosphere of the Sun, is traditionally described by magnetohydrodynamic,
fluid-like approach. Here we review some recent developments when, instead, a full kinetic description is used.
It is shown that some of the main unsolved problems of solar physics, such as coronal heating
and solar flare particle acceleration can be viewed in a new light when the small-scale, kinetic plasma description methods
are used.
\end{abstract}

\section{Introduction}

Solar corona, a tenuous and very hot outer part of the solar atmosphere which sits on top of 
other layers (photosphere, chromosphere,
and the transition region) can be seen either in white light observations 
during the solar eclipses (because of $10^6$ times more
flux of visible photospheric photons is otherwise outshining it)
or in extreme ultra-violet and X-ray observations 
from space (because of atmospheric absorption of the short wavelength radiation).
\citet{cm1919} describe the first successful observations of the green coronal line $\lambda = 5303 \AA$ 
by W. Harkness and independently by C.A. Young during the total solar eclipse of
August 7, 1869. This "new" line was attributed to a new chemical element "coronium".
It was easy to make such a mistake because just a year before, 1868,
during the solar eclipse \citet{fl1868} 
discovered a prominent yellow line $\lambda = 5880 \AA$, which could not be 
ascribed to any known chemical element at the time.
This marked a discovery of Helium after Greek word 'Helios' meaning 'Sun'. 
On Earth helium was only found about 10 years later by W. Ramsay.
The blunder with coronium went on for quite some time. Moreover, as described by
\citet{b1908}, the new gas geo-coronium extends from 200 km above Earth into the space,
and that the steady auroral arcs are due to electric luminescence in this gas.
The truth had to wait until the beginning of 1940s. In his George Darwin lecture of 
the Royal Astronomical Society \citep{e1945} entitled
"The identification of the coronal lines" Edl\'en describes the process of realising that
the mysterious coronal lines, including the green coronal line $\lambda = 5303\AA$ ,
come from so-called forbidden lines of highly ionised metals.
For example, the most intense green coronal line comes from FeXIV, iron atom with 13 electrons
stripped off. Edl\'en rightfully acknowledges that the correct identification of the
coronal lines was triggered by a letter of 13 February 1937 sent to him by W. Grotrian.
These forbidden lines are emitted from energetically unfavourable metastable
atomic energy levels. Under normal laboratory conditions electrons 
do not readily make transitions from the metastable states to lower energy states
(whilst emitting photons) due collisional de-excitation. In the extremely rarefied
conditions of the solar corona such transitions become possible.
The highly ionised metal ions can only be created in extremely hot conditions
$T >$ few $10^6$ K. Thus, at the time it was rather difficult to accept that
solar corona is so hot. Especially because this seemed in an apparent violation of
the second law of thermodynamics (in Rudolf Clausius's formulation) 
that the heat cannot flow from a cold body
(photosphere at $\approx 6000$ K) to a hotter body (corona at few $10^6$ K).
In this manner, one of the major unsolved problem in solar physics has emerged:
which physical mechanisms are responsible for heating of solar corona?
Some significant progress has been made, partcitlucalry, with the advent of
space era in solving the coronal heating problem (see e.g. chapter 9 in \citet{asch2006}).
As it happens in science, in general, 
the progress has led to yet more, previously
unknown challenges such as finding the answers to the questions: 
(i) Why solar wind speed in the polar regions 
is twice higher  during the solar minimum
than plausible solar wind models can explain?
(ii) Why the solar flares, the violent events in the corona when magnetic energy is
converted into other forms, occur on timescales much shorter than the plausible
resistive timescales predict?
(iii) Why the most of the solar flare energy is taken away by the accelerated, super-thermal
particles? and so on.
It is important to realise that reproducing or mimicking the reality in (mostly)
numerical models does not necessarily mean that we have an {\it understanding} of
physical mechanisms in action.
Here are a few examples e.g. in simple 1D coronal loop models a steady
state of few $10^6$ K plasma can be maintained by balancing
an {\it ad hoc} heat input with realistic losses such as heat conduction, radiation, etc,
or even time-varying flare dynamics can be simulated \citep{t2004}.
More realistic 3D models \citep{gn2005,g2009} mimic the solar coronal observations
and yet use unrealistically high magnetic resistivity.
No wonder if one puts sufficient heating or artificially adjusts
dissipation coefficients, a few $10^6$ K plasma can be obtained that "looks
like" the solar corona. But did we learn what generates this heat
or makes the dissipation coefficients anomalously high? Albeit,
the answer is no. Notwithstanding, something useful can be learnt
e.g. where spatially heat needs to be deposited to reproduce
the observations.
Similarly, consider the sophisticated solar wind models  
which solve numerically fluid-like equations for electrons, protons and
alpha particles \citep{li2006}. Indeed, such models can reproduce twice
as high solar wind speeds in the polar regions than near the solar 
equatorial plane, as observed during solar minima. But, this is obtained
based on an {\it ad hoc} energy flux injected into the both ion species.
Again, it comes as no surprise that {\it ad hoc} energy flux (essentially
an additional momentum added) to the ion species produces observed 
fast solar wind speeds. Again, no answer is provided to the main
question: what provides this additional energy flux (momentum)?
Is it due to absorption of waves? or due to some microscopic reconnection events
at the base of the corona?
As to the question of why the most of the solar flare energy is taken away 
by the accelerated particles, the situation is analogous.
Kinetic-scale Particle-In-Cell modelling \citep{th08} can in principle reproduce
the solar flare observations in that within one Alfv\'en time, somewhat less than half ($\leq 40$\%) of 
the initial total (roughly magnetic) energy is converted into the kinetic energy of electrons, 
and somewhat more than half ($\leq 60$\%) into kinetic energy of ions (similar to solar flare observations).
Also, a sizable fraction (up to 20\%) of the magnetic energy can be released/conversted into other
forms such as kinetic energy of super-thermal particles and to much small extent waves.
However, such simulations are performed usually in a domain sizes of several hundreds of
electron Debye length $\lambda_D=v_{te} / \omega_{pe}=2.2\times 10^{-3}$m, which is about 
$10^5-10^6$ times smaller than the solar flare particle 
acceleration site. 
Here $v_{te}=\sqrt{k_B T/m_e}$ is electron thermal speed taken at $T=10^6$K and 
$\omega_{pe}=\sqrt{n e^2/(m_e \epsilon_0)}$ is electron plasma frequency
taken at $n=10^{15}$ m$^{-3}$, commensurate to solar corona. 
The question is whether realistic behaviour mimicked in small
simulation domains will also hold if they are up-scaled to realistic sizes?
The situation is analogous to the stability of plasma in thermonuclear fusion
devices such as Tokamaks: small-scale ("laboratory"-scale) versions behave quite 
differently from the plasma stability point of view than the large,
"factory"-scale ones.

\section{Magnetohydrodynamic (MHD) vs kinetic plasma description}

The above discussion about the spatial scales, $L$, -- the characteristic
length scales of a physical system over which properties
of physical quantities (e.g. magnetic field of an active region in the
solar corona) change -- brings us to the following dichotomy or a dilemma.
One one hand, solar corona is well described by the MHD approach.
This is because if we take $L$ as the hydrostatic scale-height of the
corona (height over which pressure drops $e$-times) $\approx 50 \times 10^6$ m,
this is much larger than all relevant kinetic scales:
(i) $c/\omega_{pi} \approx 7.2$ m, ion (proton) inertial length (also called ion skin depth),
(ii) $c/\omega_{pe} \approx 0.2$ m electron inertial length (also called electron skin depth),
(iii) $r_{L,e} = v_{te}/ \omega_{ce} = 2.2\times 10^{-3}$m electron Larmor radius.
($\omega_{ce}= e B/m_e$ is the electron cyclotron frequency, which quantifies rotation of electrons
around the magnetic field on their helical path.)
On the other hand, all interesting physical processes such as 
wave dissipation, nano-scale {\it fast} reconnection (coronal plasma heating) and particle acceleration (during flares)
occur at small, kinetic scales. 
The dichotomy, depicted in Fig.~1, is in that the bulk plasma dynamics in the solar corona
(where most of the energy is stored) is observable remotely from Earth (typical 
smallest coronal structures resolved e.g. with TRACE satellite is 0.5$\arcsec$ i.e.
$\approx 0.36 \times 10^6$ m on the Sun).
But the above interesting physical processes all operate at scales below 10 m.
Thus, unless an in-situ probes are sent to the solar corona the kinetic-scales
processes will never be observed. It seems unlikely humankind will ever come
up with materials (for the probes) which will withstand the heat of the coronal
environment. However, ESA has plans for Proba mission ($http://www.esa.int/esaMI/Proba/$), in which 
the telescope of Proba-3's solar coronagraph will be mounted on one of the 
spacecraft, while the other craft is manoeuvred to accurately occult (mask) the 
main disc of the Sun. With the proposed Proba-3 arrangement, it 
is anticipated that accurate measurements will be possible from 1.05 
to 3.2 solar radii.
Note that the discussion about spatial scales $L$ can always be cast into temporal scales, $T$ or
frequencies $f=1/T$ by using characteristic speed of processes $V_{char}$ via 
$T=L/V_{char}$. For example $V_{char}$ can be the Alfv\'en speed, $V_A=B_0/\sqrt{\mu_0 n m_i}=B_0/\sqrt{\mu_0 \rho}$, 
if MHD-scale processes
are considered. This would equally be applicable to waves and reconnection. In the latter case
reconnection outflow speeds would be implied, while in the case of former -- phase speed of the wave.

\begin{figure}
 \includegraphics[scale=0.3]{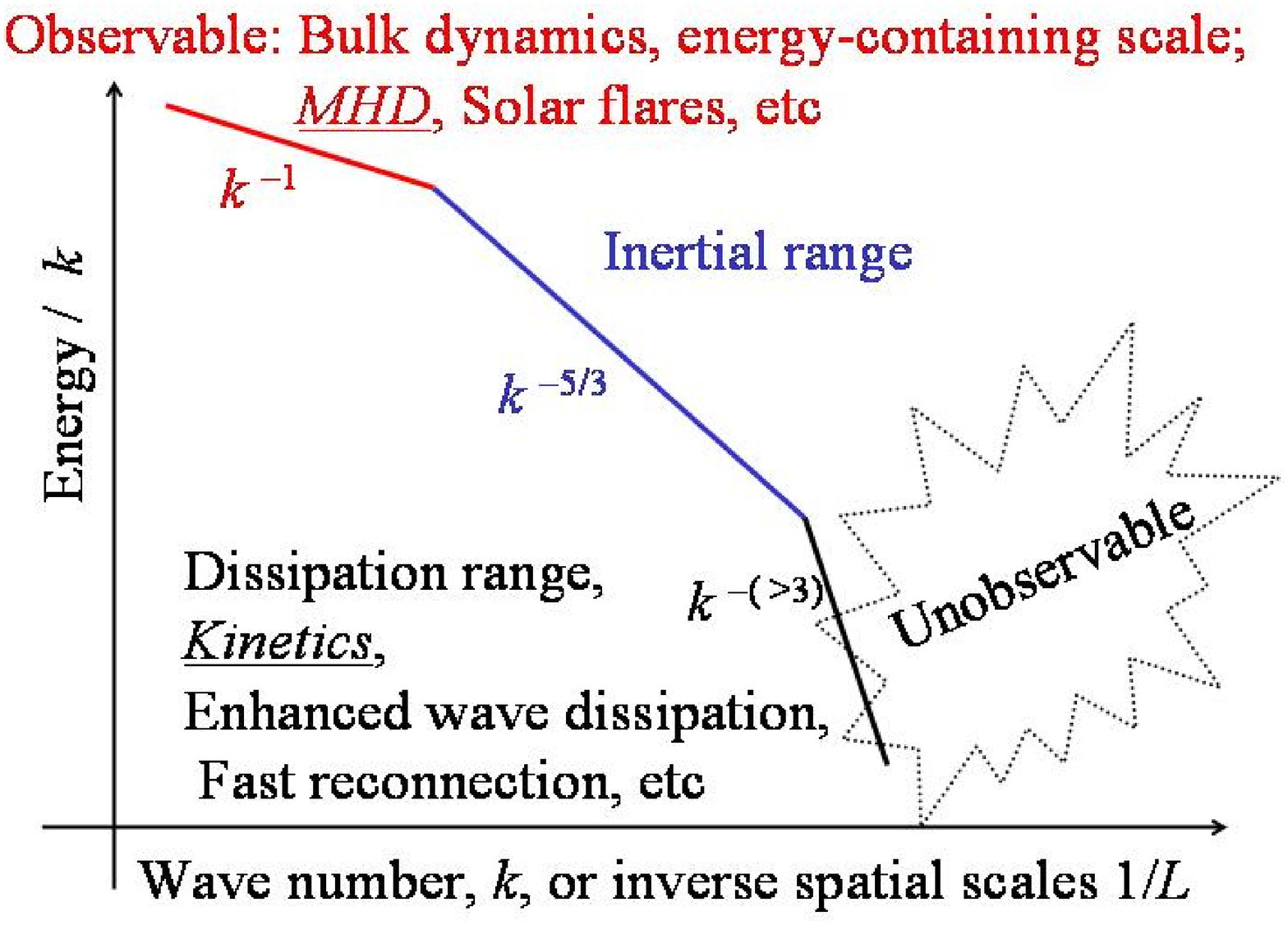}
  \caption{A sketch of typical power spectrum (energy per wavenumber $k$, (i.e. $E/k$), versus
  $k$) of magnetic fluctuations in the solar wind. Adopted from \citet{gr95}.}
\end{figure}

Let us briefly discuss conditions of when the above mentioned kinetic scales
become important. Strictly speaking MHD equations are scale-free, in a sense that
there is no physical spatial scale appearing in the equations (except $L$ which is an
{\it arbitrary} size of the system, which simply needs to stay macroscopic so that
continuum mechanical (fluid-like) description remains valid). 
The devil is in the detail, however.
In order to obtain closed set of equations one needs to express
the electric field by means of other physical quantities.
In MHD, the induction equation for the magnetic field is obtained
by plugging in $E$ into the Faraday's law (from the Maxwell's equations)
$\partial {\vec B} / \partial t = \nabla \times {\vec E}$.
Intrinsic physical scales start appearing when instead of usual MHD-version 
generalised Ohm's law \citep{ds2007}, p. 89,
\begin{equation}
{\vec E}=- {\vec v} \times {\vec B} + \eta {\vec j}
\end{equation}
one starts to use a more general form:
$$
{\vec E}=- {\vec v} \times {\vec B} + \eta {\vec j}
$$
\begin{equation}
+\frac{{\vec j} \times {\vec B}}{n e}+ \frac{m_e}{n e^2}
\left( \frac{\partial {\vec j}}{\partial t} + \nabla \cdot \left( {\vec j} {\vec v} +{\vec v} {\vec j} -\frac{{\vec j}{\vec j}}{ne}\right) \right)
-\frac{\nabla \cdot P_e}{n e}
\end{equation}
Eq.(2) essentially can be obtained from the electron equation of motion
if electron speed ${\vec v_e}$ is replaced by ${\vec v_e}= {\vec v_i} - {\vec j}/n e$.
The latter is mathematical statement of: when electron and ion speeds are different,
this causes charge separation electric field and hence current ${\vec j} = e n ( {\vec v_i} -{\vec v_e})$.
In MHD, however, ion and electron separate dynamics is ignored and ${\vec v_i}={\vec v_e}={\vec v}$.
It can be shown \citep{ds2007} that each of the non-MHD terms (second line of Eq.(2)) have 
intrinsic spatial scales associated with them: (i) the term containing ${\vec j} \times {\vec B}$ (called Hall term)
becomes important on scales comparable to $c/\omega_{pi}$, ion skin depth;
(ii) the term containing $({m_e} / {n e^2}) {\partial {\vec j}} /{\partial t} $ (called electron inertia term)
becomes important on, $c/\omega_{pe}$, electron skin depth scale; and finally
(iii) the term containing the divergence of the pressure tensor, $\nabla \cdot P_e$, (called pressure tensor term)
becomes important on scales comparable to $r_{L,e}$, electron Larmor radius.

\section{small scale effects in reconnection}
In order to better understand ordering of different spatial scales 
let us consider a simple problem of steady reconnection
when oppositely directed magnetic field lines are brought into a diffusion region by an
inflow of plasma, then they change connectivity and plasma outflow carries the
field lines away.
The change of connectivity occurs in the diffusion region
which has width $\delta$ and length $\Delta$.
In MHD, in the simplest possible formulation (Sweet-Parker model): (i) Bernoulli
equation determines the outflow speed being the Alfv\'en speed, $V_{out}=V_A$;
(ii) continuity equation prescribes in the reconnection inflow speed $V_{in}=(\delta / \Delta) V_A$;
while (iii) the generalised Ohm's law in which the advection term ${\vec v} \times {\vec B}$
is balanced by the resistive term $\eta {\vec j}$ sets the reconnection rate 
$M_{sp} = V_{in}/V_{out}=V_{in}/V_{A}=S^{-1/2} << 1$, where $S >> 1$ is the magnetic 
Lundquist number. Thus, in MHD Sweet-Parker model (in which $\Delta$ is fixed at $L$)
is producing reconnection rates much smaller what is observed in e.g. flares in solar corona.
For the parameters commensurate to solar corona
($n=1.0\times 10^{15}$ m$^{-3}$, $T=1.0 \times 10^6$ K, $L=10^5$ m, 
$B=0.01$T (100 Gauss and hence $V_A=6.9 \times 10^6$ 
m s$^{-1}$), $S=3.7 \times 10^{11}$) the classical Sweet-Parker rate is $M_{sp}=1.6
\times 10^{-6}$. 
The reconnection rate is also interpreted as the ratio of Alfv\'en
time ($\tau_A = L / V_A \approx 0.0145$ s) and
resistive (or reconnection) times.
This means in the Sweet-Parker
model resistive (or reconnection) time is $0.0145 / S^{-1/2}$ s $= 0.1$ days.
On contrary, in the observations flares last for up to tens of minutes.
Petschek model alleviates this problem by making the diffusion region length, $\Delta << L$.
This means that the plasma inflow speed $V_{in}=(\delta / \Delta) V_A$ a sizable fraction
of the Alfv\'en speed making reconnection rates faster and hence flare times shorter
commensurate to the observations. However, Petschek model is a phenomenological one
(not derived from the first principles) and for it to work (as demonstrated by
numerous numerical simulations) it requires plasma resistivity to be
non-uniform and be enhanced on the diffusion region (so-called anomalous resistivity).
Whilst, the latter being quite plausible, given that turbulent transport
is known to enhance the dissipation/resistive coefficients dramatically,
we have not sent in-situ probes to the solar corona yet, and hence we do not
know for certain whether it is turbulent or not!

It should be noted that in the laboratory plasma Magnetic reconnection experiment (MRX)
measurements show indication of the reconnection rates or equivalently
anomalous resistivity values 10s and even 100 times larger that Sweet-Parker
model predicts \citep{yamada}. The MRX experiment plasma {\it is not in the turbulent}
state. What is observed, however, is that the  reconnection rates or equivalently
anomalous resistivity values are attained when the transition from collisional (resistive MHD)
to collisionless (kinetic) regime takes place.

\begin{figure}
 \includegraphics[scale=0.3]{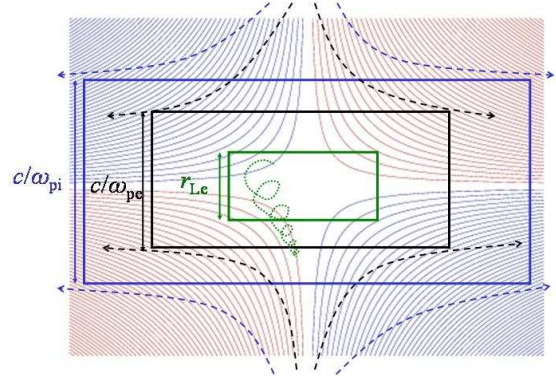}
  \caption{A sketch of kinetic-scale magnetic reconnection.
  Blue and red curves in the quadrants show an opposite sign magnetic flux.
  The outer, blue box shows a region where ions become demagnetised, whilst
  electrons are still magnetised. From this scale ($c/\omega_{pi}$) and below 
  the magnetic field is carried by electrons. Black box with width $c/\omega_{pe}$
  shows electron diffusion region where magnetic connectivity change take place. 
  The green box with width $r_{L,e}$ shows region where electron meandering motion 
  makes pressure to become a tensor with off-diagonal (non-gyrotropic) terms playing
  a major role. Blue and black dashed lines with arrow heads show ion and electron flow,
  respectively. Note how ion flow deflects at the scale $c/\omega_{pi}$,
  while electron flow does so at $c/\omega_{pe}$. The green wiggle in the centre
  show an electron orbit. Its radius is small and follows a magnetic field line
  away from the diffusion region. The radius becomes large and motion chaotic (meandering)
  as the magnetic field drops to zero close to the magnetic null.}
\end{figure}

So, is there an alternative way (i) to have fast reconnection i.e. get solar flare times right?
(ii) also to demonstrate that the large proportion of flare energy goes into the accelerated particles?
and (iii) not to make assumptions about the turbulent state of the corona (hence postulating the
anomalous resistivity)? It turns out that plasma {\it kinetic} approach 
can provide answer "cautious yes" to all of these fundamental questions. 
Here is how: For the solar coronal
parameters: a temperature of $1.0 \times 10^6$ K, Coulomb logarithm of 18.1, the 
Lundquist number (using Spitzer resistivity) is 
$3.7 \times 10^{11}$. Here $L=10^5$ m was used.
One of the reasons for going beyond resistive MHD is comparing
typical width of a Sweet-Parker current 
sheet $\delta_{sp}=S^{-1/2} L=0.16$ m to the ion skin depth.
Typical scale associated with the Hall term in the generalised Ohm's law at which deviation from
electron-ion coupled dynamics is observed is, $c/\omega_{pi} =  7.2$ m. Here 
particle number density of $n=1.0 \times 10^{15}$ m$^{-3}$ has been  used.
Hence, the fact that $ c/(\omega_{pi} \delta_{sp})=44 \gg 1 $ justifies
going {\it beyond single fluid resistive MHD approximation} (a 
similar conclusion is reached by \citet{yamada} in their Fig. 12).
In other words, ion skin depth scale is reached first
before resistive MHD Sweet-Parker current 
sheet. 
Let us put this fact (that $ c/(\omega_{pi} \delta_{sp}) \gg 1 $)
in the visual context using Fig.~2:
On a global (bulk MHD) scale $L$ electrons and ions inflow towards the diffusion region
(where magnetic field connectivity changes) in a coordinated way as if they are glued to each other.
When the flow approaches ion skin depth scale $c/\omega_{pi}$ ions become demagnetised
(magnetic field is no longer is frozen into ion fluid) and magnetic field
is carried forward by electrons. At $c/\omega_{pi}$ scale the Hall term in the generalised Ohm's law dominates.
This picture is broadly corroborated by kinetic, particle-in-cell simulations of
x-point collapse \citep{th07}. In panel (a) of Fig.3 we see that for electrons
the diffusion region width is indeed $c/\omega_{pe}$, while for ions, panel (b) of Fig.3,
the width of the region where ions start to deflect from the current sheet
(region where ions become demagnetised) is about $c/\omega_{pi} =10 c/ \omega_{pe}$
(note that here the ion to electron mass ratio of $m_i/m_e=100$ has been used).
This continues until electron skin depth scale $c/\omega_{pe}$ is reached where
electron inertia term takes over.
Further down the spatial scales, when electron Larmor radius scale $r_{L,e}$
is reached electron meandering motion (dashed green wiggled curve in Fig.2) sets in
\citep{hs1997}. This makes the non-gyrotropic (off-diagonal) 
components of the electron pressure tensor to dominate other terms in the
generalised Ohm's law, and hence makes them responsible for breaking the frozen-in condition. 
The concept of pressure {\it tensor} is somewhat counter-intuitive.
This can be comprehended as follows: away from the electron diffusion region
pressure is scalar (isotropic) and electrons flow along (are frozen into) the magnetic field lines.
Inside the the diffusion region $\leq r_{L,e}$ electrons become demagnetised and
execute meandering (rather chaotic) motions. These are such that pressure becomes
different as one travels in different directions (i.e. pressure becomes a tensor
as opposed to a scalar).
Previous results on collisionless reconnection both in tearing unstable 
Harris current sheet \citep{kuz98,hesse99,birn01,pritchett01} and stressed X-point collapse 
\citep{th07,th08} have shown that magnetic field is frozen into electron fluid
and the term in the generalised Ohm's law 
that is responsible for breaking the frozen-in condition is
electron pressure tensor off-diagonal (non-gyrotropic) component gradients. 
Thus, there is a case for inclusion of the
{\it electron pressure tensor non-gyrotropic components} in a 
model of collisionless reconnection.
Analytically, this was achieved by \citep{t08}.
In this simple model instead of balancing the advection term in the
generalised Ohm's law with the resistive term (as discussed above),  non-gyrotropic
components of the electron pressure tensor gradient were used.
This produced simple analytical results which explain well
(i) the reconnection rate and (ii) width of the electron diffusion
region measurements in MRX experiment.
\begin{figure}
 \includegraphics[scale=0.4,angle=90]{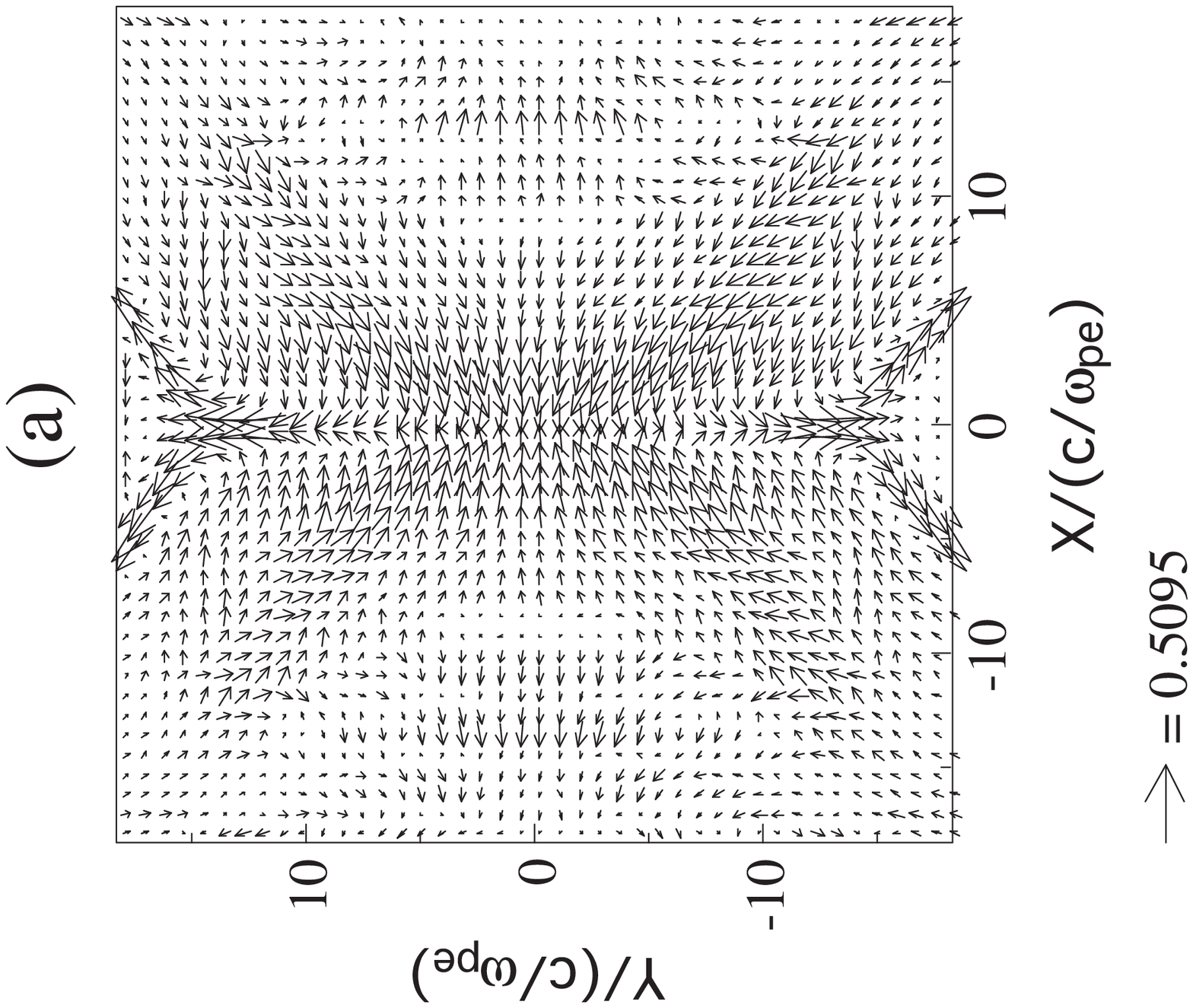}
 \includegraphics[scale=0.4,angle=90]{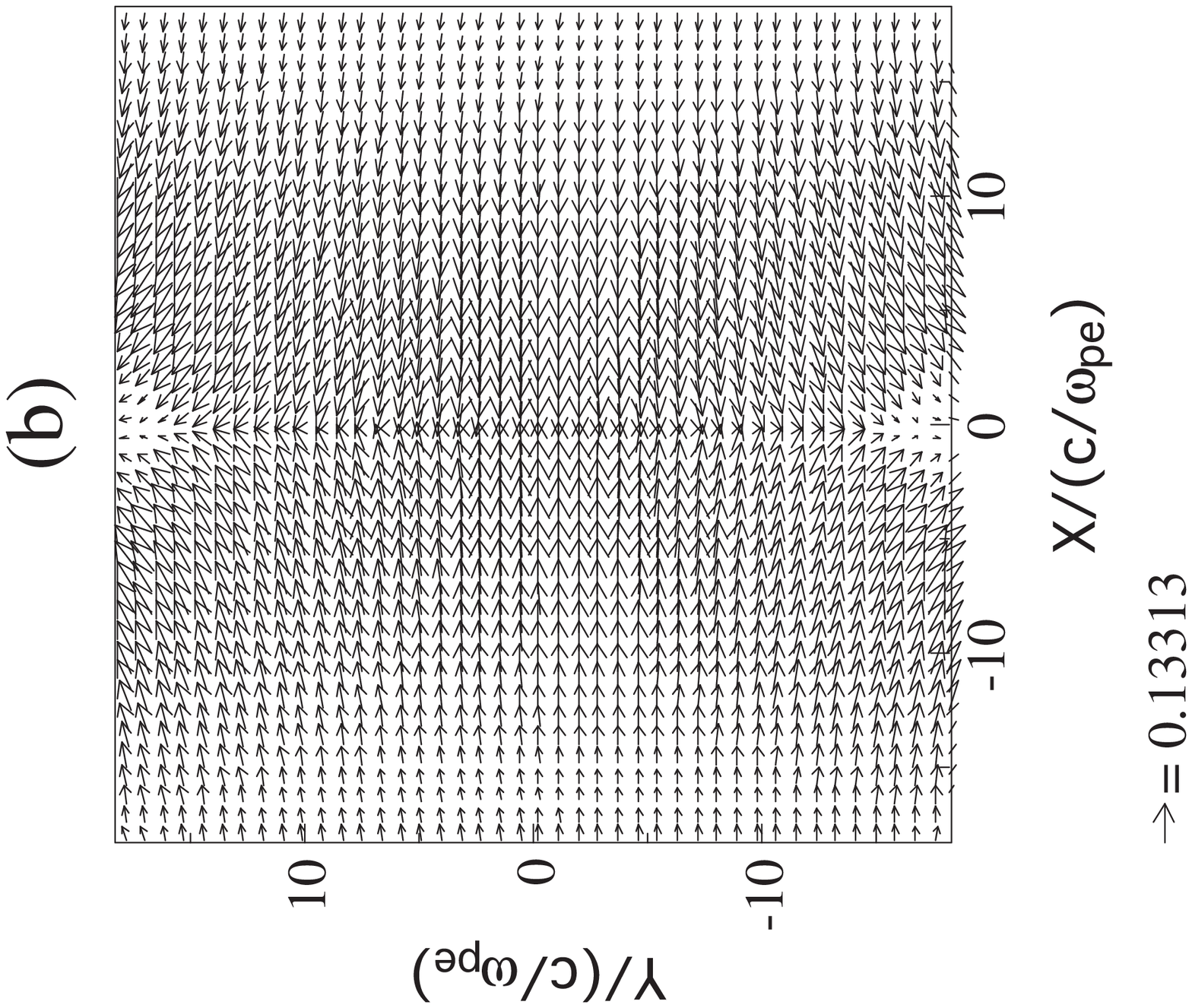}
  \caption{Electron (a) and ion (b) flow patterns at the peak of 
  time-transient reconnection of x-point collapse. Adapted from \citet{th07}, who used
  particle-in-cell, kinetic approach for the numerical simulation.
  The arrow size represents instantaneous, cell-averaged 
  velocities normalised by speed of light.
  Note that for electrons the diffusion region width is about $c/\omega_{pe}$, while
  ion flow deflects from the diffusion region at a width $c/\omega_{pi}$ (here $c/\omega_{pi}=10 c/\omega_{pe} $), 
  as sketched in Fig.~2.}
\end{figure}

There is a growing amount of work \citep{ds2007,th07,th08} that 
suggests that in the collisionless regime,
on the scales less than $c/\omega_{pi}$ magnetic field is frozen into the 
electron fluid rather than bulk of plasma.
One can write in general $\vec v_e=\vec v_i- \vec j / (e n)$. This relation clearly
shows that in collisional regime (when the number density $n$ is large), the
difference between electron and ions speeds diminishes $v_e=v_i=v$. However, as one enters
collisionless regime (when the number density $n$ is small) the deviation between
electron and ion speeds starts to show.
In \citet{th08}
we  proposed a possible explanation why the reconnection is fast when
the Hall term is included. Inclusion of the latter means that
in the reconnection inflow magnetic field is frozen into {\it electron} fluid.
As it was previously shown in \citet{th07} (see their Figs.(7) and (11))
speed of electrons, during the reconnection peak time, is
at least 4-5 times greater than that of ions. This means that electrons can
bring in / take out the magnetic field attached to them into / away from 
the diffusion region
much faster than in the case of single fluid MHD which 
does not distinguish between
electron-ion dynamics. Thus, it is clear that inclusion {\it magnetic field transport by electrons}
is crucial in resolving of above mentioned problems MHD description
of the solar corona faces.

To summarise, some of the time-dependent collisionless reconnection models such as
tearing unstable 
Harris current sheet \citep{kuz98,hesse99,birn01,pritchett01} and stressed X-point collapse 
\citep{th07,th08}
where able: (i) to show that a sizable fraction of the magnetic energy
can be converted/released into heat and accelerated, super-thermal
particle energy (see e.g. Fig.13 from \citet{th07});
(ii) obtain fast reconnection rates and hence magnetic energy release rate
commensurate to solar flare observations (see e.g. Figs.3 and 9 from \citet{th07});
(iii) reproduce the observational fact that the large proportion of flare energy 
goes into the accelerated particles in the correct
partition, i.e. somewhat less than half (40\%) of the initial 
total (roughly magnetic) energy is converted into the kinetic 
energy of electrons, and somewhat more than half (60\%) into kinetic energy of ions
(see e.g. Figs.3 and 4 from \citet{th08}); 
(iv) not to invoke anomalous resistivity.
However, despite this success there are main challenges ahead:
(i) the processes described by the collisionless reconnection models are small scale ($<c/\omega_{pi} \approx 10 $ m).
Indeed, on scale less than ion skin depth magnetic field is advected into the diffusion 
region by electrons and since electrons are lighter and move faster, the reconnection
is accordingly fast. But, sufficiently far away from the electron diffusion region,
electron flow must slow down to bulk plasma speeds when MHD description 
takes over (ions and electrons become glued to each other), where and how this happens?
Recent PIC simulations \cite{do06,sh07} suggest that the electron 
diffusion region length can extend for much longer distances 
downstream than previously thought. Why does this happen is not understood.
(ii) How this small-scale dynamics on the kinetic scale
translate into MHD scales? i.e. despite producing fast reconnection rates,
these processes happen in physically small volumes.
Will they affect MHD-scale (say a $L=10^6$ m) processes?
It will be impossible to find out until we can perform a numerical simulation
$10^6$m$\times 10^6$m$\times 10^6$m using realistic coronal
active region fields as an input for Particle-In-Cell or Vlasov
simulation. The latter would seem many decades away even if Moore's law
(that the number of transistors that can be placed inexpensively on an 
integrated circuit has increased exponentially, doubling approximately every two years)
continues to hold in the future. 
Until such times, we can do $10$m$\times 10$m$\times 10$m
kinetic simulation and hope for the best that our models will 
scale correctly towards the larger (physically meaningful) spatial scales.

\section{small scale effects in waves}
Study of MHD waves in the solar corona has seen steady progress over last two
decades. There are two aspects to the study:
(i) waves carry information about the medium they propagate in.
This is used in the field of solar coronal seismology
to infer physical parameters of the corona (see  for reviews \citet{nv05,dem09}).
For example traditional method of measuring magnetic field
in solar photosphere is based on the Zeeman effect.
However, in solar corona plasma is too dilute rendering this method
unusable. Based on the seminal theoretical work of \citet{er83}, 
transverse oscillations of solar coronal EUV loops were used to successfully measure 
the magnetic field \citep{no01}.
(ii) waves transport mechanical (and magnetic) energy and if
dissipated may produce plasma heating. 

In the MHD approximation, in a uniform plasma penetrated by a 
uniform magnetic field, $\vec B_0$, there
are three distinct types of waves: fast and slow magnetosonic and Alfv\'en waves.

Fast magnetosonic waves are compressible MHD waves which have an advantage
of propagating across the magnetic field. This implies that the heat
produced by the wave dissipation can spread across and cover a large area.
The fast magnetosonic waves that are generated in the solar interior
cannot reach the corona because of the reflection on the pressure/density
gradient (due to gravitational stratification). Thus, initially they
were excluded from coronal heating models. However,
in the transversely inhomogeneous plasma  there are
two possibilities to couple Alfv\'en waves (which have no problem of 
crossing the gradient) with fast magnetosonic waves: either by
non-linear coupling in 2.5 MHD approximation \citep{nrm97,t2001}; or
linear (much more efficient) coupling in 3D MHD approximation \citep{tn2002,tnr2003}.

Slow magnetosonic waves are also compressible, but propagate preferentially along
the magnetic field. Normally, these waves are discounted from coronal heating candidate
list on the grounds of not carrying (less by four orders of magnitude) sufficient flux \cite{dem09}.
However, the latter estimate is based on the assumption of just a single harmonic (fixed frequency) and
\citet{tn2001} have demonstrated
 that if a wide spectrum (continuous spectrum with a plausible frequency (or wavenumber) range)
of slow magnetosonic waves exists in the corona, than their dissipation would provide
sufficient heating. Whether such spectrum exists is an open question. However since in the
nature there is always a cascade of energy from large scales to small scales (see e.g. Fig. 1),
this wide spectrum conjecture seems plausible.

Alfv\'en wave is an incompressible, transverse wave (such as e.g. electromagnetic wave), 
with frequency, $\omega_A$,
much smaller that ion cyclotron frequency, $ \omega_{ci}=e B/ m_i$.
In an Alfv\'en wave the background magnetic field tension
provides restoring force, while  plasma ions provide inertia
(the two key ingredients for any oscillatory motion).
Alfv\'en waves are easy to excite, e.g. a sheared plasma flow
across magnetic field. However, they are notoriously difficult
to dissipate because of small resistivity of the solar corona.
i.e. Alfv\'en wave is a good vehicle for transporting energy (e.g.
from convective motions below the photosphere into the corona), but this
"vehicle" has virtually no breaks (hence cannot be stopped!). Thus, 
in order to deposit energy in the first  $\approx 50 \times 10^6$ m
of the corona (the hydrostatic scale-height) above the transition region,
some enhanced dissipation mechanisms where proposed in the past.
In the solar corona,  basic magnetic structure that is encountered is either 
a closed loop  (during 11-year cycle minimum 
more of these are found near solar equator and low latitudes); or open
magnetic strictures such a plumes (during the solar minimum these are mostly located near the 
north and  south polar regions). The common feature of these coronal 
structures is the density variation {\it across } the magnetic field.
i.e. inside the loops and plumes density is enhanced by 3 -- 10 times compared
to the surrounding plasma. In the case of closed magnetic structures this
enhanced density is believed to be due to material evaporation from
denser layers of the sun such as transition region and chromosphere.
In the open magnetic structures the density enhancement is either 
due to a solar wind and again material evaporation from below (which could be
either due to reconnection or any other process which increases plasma temperature
so that the balance no longer holds).
There are two mechanisms for the above mentioned enhanced Alfv\'en wave
dissipation and absorption in the solar corona: phase mixing \citep{hp83} and resonant absorption \cite{ion78}.
In both mechanisms transverse inhomogeneity of plasma (i.e. plasma with density variation across
the uniform magnetic field) plays a crucial role (albeit for different reasons). 
Brief review of both topics can be found in \citet{dem08}.

\begin{figure}
 \includegraphics[scale=0.3]{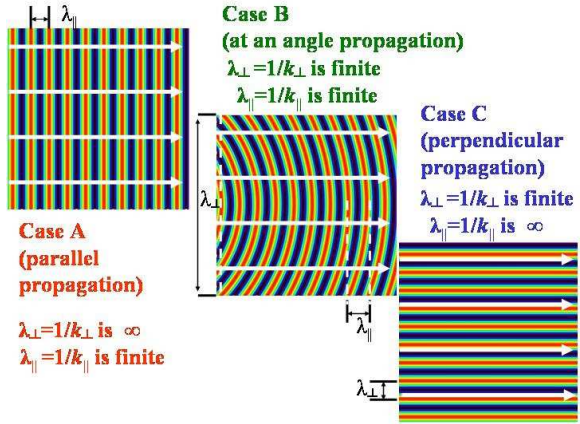}
  \caption{A sketch of parallel, at an angle and perpendicular propagation of a wave.}
\end{figure}

For particle acceleration and efficient {\it non-resonant} 
wave-particle interactions (e.g. converting wave energy to particle kinetic energy and heat if inter-particle 
collisions are efficient), it is necessary for a wave in question to have magnetic field-aligned electric field.
The word non-resonant is crucial here because, interaction of waves with particles when
(i) the wave phase speed $V_{ph}=\omega / k$ coincides with particle (thermal) speed $v_{th,\alpha}=\sqrt{k_B T / m_\alpha}$ 
leads to the Landau resonance; or
(ii) to the  cyclotron resonance, when the rotation frequency of the  electric field vector of the circularly polarised
wave, $\omega$ coincides with the particle's cyclotron frequency, $\omega_{c \alpha}= eB/m_\alpha$ (and when both the
wave electric field vector and particle rotate in the same direction).
In the both cases particle experiences almost constant electric field, which facilitates
efficient wave-particle resonant interaction.
Away from the resonance $V_{ph} \gg$ or $\ll v_{th,\alpha}$ in the case of Landau damping;
or $\omega \gg$ or $\ll \omega_{c \alpha}$ in the case of cyclotron resonance, wave electric field 
oscillates too fast or too slow and has no effect on plasma particles of species $\alpha$.
In the MHD approximation Alfv\'en wave propagates strictly along the magnetic field (with some finite 
parallel to the magnetic field wave number $k_\parallel \not =0$)
as shown in case A, Fig. 4. In this case the perpendicular wave number $k_\perp=0$.
For the field aligned electric field to exist $k_\perp \not=0$ condition should be true.
Case C, Fig.4 also shows why the fast magnetosonic waves are believed to be efficient particle
accelerators. This because they preferentially propagate across the magnetic field and hence
posses finite $k_\perp \not=0$.
For Alfv\'enic waves $k_\perp \not=0$  (see case B, Fig. 6) only possible when kinetic-scale 
effects are included. 
In particular, it has been realised \citep{hc76} that
when externally applied wave frequency becomes resonant with the Alfv\'en frequency
$\omega_A = k_\parallel V_A$, a resonance occurs and the driving wave converts
into kinetic Alfv\'en wave (KAW) which has the finite field-aligned electric field.
More generally, generation of the parallel (magnetic field-aligned) electric field
occurs when the applied low-frequency ($\omega < \omega_{ci}$) Alfv\'en wave has
$k_\perp$ comparable to any of the {\it kinetic spatial scales}.
Such Alfv\'en waves (with $E_\parallel \not =0$) are called {\it dispersive Alfv\'en waves} (DAW) \citep{stas00}.
The latter split into two classes: kinetic Alfv\'en waves (KAW) and inertial Alfv\'en waves (IAW).
The distinction between the two is drawn according to the value of plasma beta,
which is the ratio of thermal and magnetic pressures, i.e. $\beta=p/(B^2/2\mu_0)$. Plasma beta can also be related
to the ratio of the sound speed and Alfv\'en speed squared, i.e. $\beta =(2/\gamma)(c_s^2/V_A^2)\approx c_s^2/V_A^2$, 
the ratio of specific heats $\gamma$ in e.g. adiabatic case is 5/3. Here $c_s=\sqrt{\gamma p/\rho}$ is the sound speed.
Physically, plasma beta prescribes importance of either thermal effects or magnetic field.
E.g in solar corona $\beta \ll 1$ meaning that magnetic fields play a dominant role
in coronal plasma behaviour. In the context of dispersive Alfv\'en 
waves plasma beta prescribes which term in the generalised Ohm's law supports 
parallel electric field \citep{stas00}.
If $\beta < m_e/m_i$, i.e. $v_{te}, v_{ti} < v_A$ then the dominant mechanism for generation 
of parallel electric field in the generalised Ohm's law is the electron inertia and 
one has inertial Alfv\'en waves (IAW).
If $\beta > m_e/m_i$, i.e. $v_{te}, v_{ti} > v_A$ then the dominant mechanism for generation 
of parallel electric field in the generalised Ohm's law is electron pressure gradient 
and one has kinetic Alfv\'en waves (KAW).

In the solar corona, particle acceleration e.g. during solar flares
is generally believed to be occurring as a result of magnetic reconnection.
For description of particle dynamics either test particle approach \citep{zhg05,db08}
which is ignores self consistent fields or more recently
kinetic, particle-in-cell simulation \citep{szh09} has been used.
The problem with particle acceleration during solar flares is associated with
the numbers of particles involved.
In the standard flare model, reconnection event occurs higher up the corona
and as result 50-80 \% of the flare energy is converted into accelerated particles (electrons and ions).
These rush down towards denser layers of the solar atmosphere guided by coronal magnetic field
structures and produce
X-ray emission via bremsstrahlung. The number of detected X-ray photons can be related
(via certain model assumptions) to the number of accelerated electrons.
For a typical flare more than $10^{39}$ electrons
are accelerated, yielding a huge current sheet (with a volume of $10^{28}$ cm$^3$) 
that must remain stable for more than 60 seconds. This is an unlikely course of events because of
the dissolution (thinning) of the current sheet \citep{bm77}.
In effect, the above numbers imply that all of the electrons in the
current sheet need to be evacuated.
Another problem is with the return currents. A beam of electrons
injected into plasma is known to be compensated by the generated 
counter beam. Thus, expecting a bunch of
all $10^{39}$ electrons happily travels towards the footpoints
of the magnetic structure is naive.
A viable alternative for transporting flare energy by the 
beam of accelerated particles, is waves.
As we know waves can transport energy without mass transport.
Therefore, if flare energy is delivered to footpoints by waves
then the above mentioned difficulties (the number problem and return
currents can be avoided).
In this scenario, reconnection flare event produces dispersive Alfv\'en 
waves (which posses parallel electric fields). These subsequently travel to the
footpoints, then accelerate particles in the vicinity of footpoints.
The accelerated particles (mostly electrons, because these waves are known to preferentially
accelerate electrons due to their small inertia, whilst producing ion heating -- broadening
of their distribution function), in turn,  produce the observed X-rays via the
bremsstrahlung. Using particle-in-cell simulation,
\citet{tss05} have recently explored how circularly polarised Alfv\'en wave with 
$\omega=0.3\omega_{ci}$ propagates in
a transversely inhomogeneous plasma. 
This frequency is somewhat higher than that for
MHD wave and yet smaller that ion cyclotron resonance. Such wave 
strictly speaking is called Ion cyclotron rather than Alfv\'en wave. In the $\omega \ll \omega_{ci}$
limit, however, is behaves like an Alfv\'en wave.
The density in the middle of simulation domain
was smoothly increased by a factor of 4 (a ramp-like function $\rho(y)=1+3\exp\left[
-((y-100)/50)^6\right]$) across the uniform 
magnetic field in order to model a coronal loop (accordingly
temperature was varied as inverse of density so that total pressure balance is preserved).
The widths of the transverse inhomogeneity on each side of the ramp is $\approx c/\omega_{pi}$.
\begin{figure*}
 \includegraphics[scale=0.6]{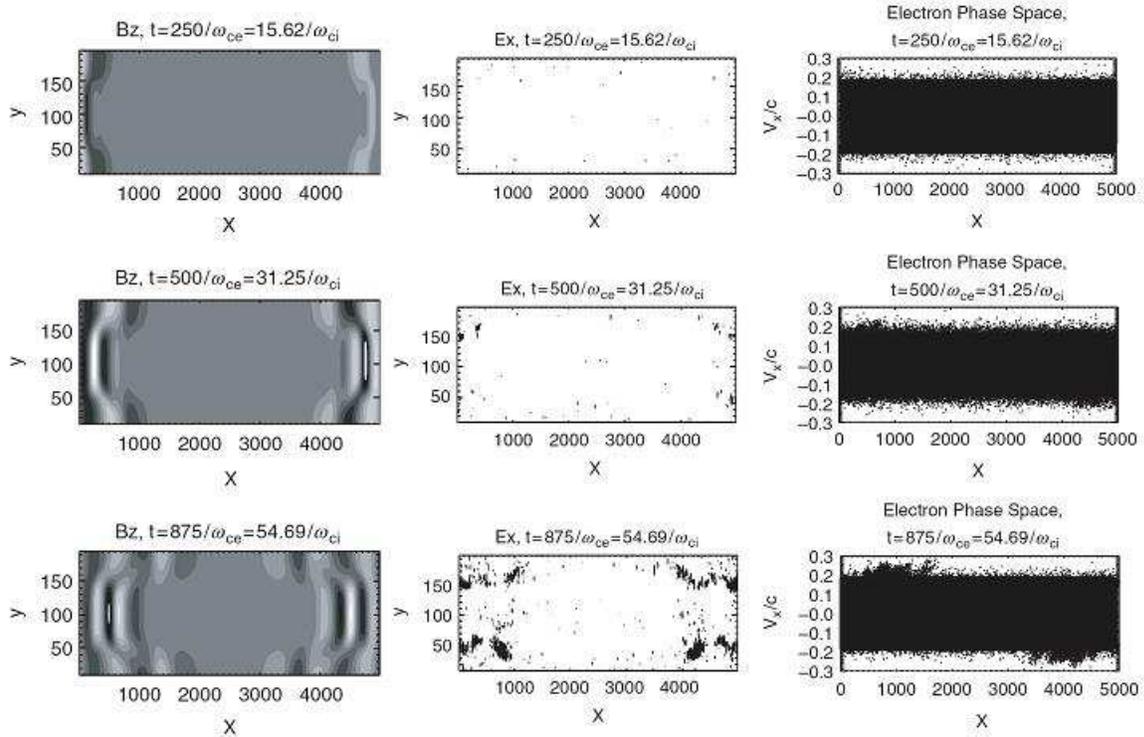}
  \caption{The left column shows three snapshots of Alfv\'en wave magnetic field component.
  The middle columns shows the generated parallel electric field, $E_{\parallel}$.
  The right column shows parallel velocity phase space of electrons.
  See text for details.}
\end{figure*}
Fig.~5's left column demonstrates that initially plane Alfv\'en wave front becomes distorted
(middle part travels slower)  because
its phase speed ($V_A(y)=B_0/\sqrt{\mu_0 \rho(y)}$) is a function of
density, which in turn is a function of a coordinate $y$  across the uniform
magnetic field applied along $x$-axis. It  is also interesting 
to note that in the strongest density gradient regions $y=50$ and 150
the propagating Alfv\'en wave damps. In the middle column of Fig.5
parallel electric field $E_\parallel=E_x$ is shown.
At time $t=0$, $E_\parallel$ is zero everywhere, but as the Alfv\'en wave propagates
and its wave front becomes distorted $E_\parallel$ starts to grow.
It typically attains values of few $10^{-3} (m_e c \omega_{pe}/e)$.
For solar coronal conditions this is typically exceeds the
Dreicer electric field  (which is associated with the particle
acceleration runaway regime \citep{dre}) $10^6$ times!
Hence, the generated parallel electric field efficiently accelerates electrons,
as can be seen in the right column of Fig.5.
The latter shows parallel to the magnetic field phase space
($V_x$ vs. $x$) where each dot corresponds to an electron (total of $\approx 5 \times 10^8$).
We gather that in the regions where $E_\parallel$ is generated, the
number of particles with increased field-aligned velocity $V_x$ is increased
(i.e. particles are accelerated).
Note that this coincides with the regions where Alfv\'en wave enhanced damping occurs.
\citet{th08pm} performed a parametric study, exploring how the
new mechanism of parallel electric field generation and acceleration 
of electrons discovered  by \citet{tss05}
depends on problem parameters, such as the variation of 
frequency and amplitude of the applied (driving) Alfv\'en wave, as well as
plasma beta, affect levels  attained by the $E_\parallel$  and the fraction 
of accelerated particles (the latter is defined as percentage of electrons 
with speeds above the electron thermal speed in the density gradient regions).
It was established that (i)  $E_\parallel$ is always orders of magnitudes greater than
Dreicer electric field and (ii) the faction of accelerated electrons ranges
within 20-50\%. The latter makes the discovered electron acceleration mechanism quite efficient
and potentially capable of resolving the above mentioned
problems of standard flare models which use reconnection-produced electron beams instead of
waves as the means of delivering energy flux to the footpoints in order to produce
the X-rays.
It should be noted that the transverse inhomogeneity is crucial in the model of
\citet{tss05} (see also further analysis paper \citep{t07}). In fact, when the transverse density inhomogeneity is removed,
i.e.  when Alfv\'en wave front remains always perpendicular to the uniform
background magnetic field, no parallel electric field generation or
electron acceleration is observed (see Figs.7-10 in \citet{tss05}).
In terms of the above $k_\perp$ discussion, this can be explained as following:
When the density across the magnetic field is inhomogeneous,
despite the fact that initially $k_\perp=0$ as the Alfv\'en wave front
starts to deform $k_\perp$ is efficiently generated.
When on the other hand density is constant everywhere
then Alfv\'en wave front is always at right angles to the magnetic field, i.e.
Alfv\'en wave propagates strictly along the field, and thus $k_\perp=0$  at all times.
Recently, \citet{fh08} revisited the problem of solar flare
electron acceleration with Alfv\'enic pulses. Further, \citet{mf09} quantified
the full details, namely, exploring how fraction of accelerated particles
is affected by such crucial parameter as transverse length-scale 
$l_\perp \simeq 2\pi /k_\perp$ of the Alfv\'enic pulse. They found that significant fraction of
electrons would be accelerated if $l_\perp$ is few meters or less.
A clear distinction should be made between models of \citet{tss05} and \citet{mf09}.
As explained above, in the case of former, $k_\perp$ is {\it initially zero}
and it seems that such wave is more plausible to exist in nature.
This is because it represents simple, ubiquitous Alfv\'enic {\it harmonic} wave.
Recall that even in MHD approximation $k_\perp$ is also zero at all times.
It is crucial in \citet{tss05}'s model that plasma has density inhomogeneity
mimicking transverse density structure of the coronal loop.
It is this inhomogeneity that makes Alfv\'en wave front to distort and
create finite $k_\perp$ and hence generate $E_\parallel$.
In the case of \citet{mf09}'s model $k_\perp \not=0$ (and hence $E_\parallel \not =0$) {\it from the start}
and  there is no transverse density inhomogeneity. Also, a pulse instead of harmonic have is
considered.
If we try to compare plausibility of the two models the following considerations
come to mind. Both models "work" i.e. generate large enough
$E_\parallel$ and accelerate sufficient proportion of electrons if the following
are true:
In \citet{tss05}'s model a large scale Alfv\'enic, harmonic wave initially has $k_\perp=0$, interacts with the
transverse density inhomogeneity which has a scale of $l_{inhom}\simeq c/\omega_{pi} \simeq c/\omega_{ci}$.
In \citet{mf09}'s model (see their Eq.(28)) 
Alfv\'enic pulse with $k_\perp \simeq 1 /[ A (c/\omega_{pi})]$
travels in the homogeneous plasma (where $A$ is the perturbation amplitude). 
In effect, both models imply either small scale transverse inhomogeneity
of the medium or small scale transverse wave number. It is unclear at this stage
whether we have such small scale phenomena in solar corona.
Perhaps, in situ coronal probes or remote sensing from distances much small than Sun-Earth distance
could shed some light, as this is currently possible in Earth magnetospheric applications.
However, until such times the jury on this topic will remain out.

In summary, wave based models offer attractive solutions to the above mentioned 
outstanding problems  in solar flare particle acceleration.
It is yet to be clarified however, how small-scale (kinetic-scale) phenomena feeds into the large-scale 
(MHD-scale) phenomena. From the fundamental point of view, of course kinetic-scale modelling
includes all the essential physics. However, realistically shortcomings of man-made even the
largest parallel supercomputers will unlikely offer us the performance we need.
E.g. a full 3D Vlasov code $128$ grids in the 6-dimensional phase space 
(3 spatial dimensions and 3 velocity dimensions) require 32 Terabytes (32,000 Gb) of RAM.
Given that spatial grid size is the Debye length ($\lambda_D=v_{th,e}/\omega_{pe}=2.2\times 10^{-3}$ m in the
solar corona), in 3D this offers rather uninterestingly small physical volume.
Thus, there is always a restriction imposed to consider a lower dimensional physical systems.

\section{Challenges ahead}

Despite significant progress in understanding of physical processes ongoing in solar atmosphere --
it is enough to take a look as NASA or ESA webpages describing a number of previous and existing successful
space missions, as well future mission plans -- there is still a long and exciting way to go!
As far as coronal heating problem is concerned, in author's opinion
further progress could be made if following considerations are noted.

From the observational perspective it would be rather important to investigate
heating of active regions (ARs). AR is a volume of solar atmosphere above sunspots.
They are important because during solar maximum ARs can be responsible for $\approx 80$\%
of heating of the entire corona \citep{asch07}.
Top left of Fig.6 shows TRACE 171 $\AA$ 
image of an AR. The box in Fig.6 shows potential
extrapolation (using Green's function method) of magnetic field 
lines based on an input from the measured normal component
of the photospheric magnetic field that can be used for numerical modelling.
Ideally, if we have {\it simultaneous}, high time cadence
 2D imaging and 2D Doppler shift data at several
wavelength, $\lambda$, corresponding to emission from e.g.
white light (continuum, photosphere, $\log(T)=3.7$), 
HeII ($\lambda=304$\AA, chromosphere, transition region, $\log(T)=4.7$),
FeIX ($\lambda=171$\AA, quiet corona, upper transition region, $\log(T)=5.8$)
FeXII, FeXXIV ($\lambda=193$\AA, corona and hot flare plasma, $\log(T)=6.1, 7.3$), etc
then one would able to Fourier transform the data to see e.g.
how much wave power spectrum is lost (dissipated) from one height 
(a horizontal cross-section through the box in Fig.6) to another.
Essentially, different line forming ions probe different temperatures and
therefore different heights in the box in Fig.6.
This would ultimately enable us 
to make a judgement to what extent {\it waves} contribute to the
AR and coronal heating? 
Also, measuring the abundance of the observed bi-polar jets, which are believed to be signatures of
small-scale reconnection, could enable us to make a judgement
on {\it reconnection}'s contribution to the coronal heating.
NASA's
Atmospheric Imaging Assembly (AIA) on board of the Solar Dynamics Observatory (SDO), 
to be launched in November 2009, will make such unprecedented progress.
However, AIA's resolution of about 1 arcsec and time cadence of 10 seconds or better
will naturally limit the types of waves it can detect.
Perhaps a better job could be done by a ground based project called
Rapid Oscillations in Solar Atmosphere (ROSA), $http://star.pst.qub.ac.uk/rosa/$,
which will have 0.1 arcsec resolution and will take in 30-125 frames per second!
However, ROSA being a ground based instrument, 
can only probe photosphere/chromosphere/TR lines, i.e. cannot probe corona, and
yet a whole wealth of new data is expected which will hopefully lead to new
discoveries.
On a negative note, with SDO's AIA and ROSA importance of kinetic effects still cannot be probed, because
1 arcsec on the Sun is $0.72 \times 10^6$m -- orders of magnitudes larger than any
kinetic scale discussed above.
The concept of measuring wave dissipation with height has been used recently 
by \citet{ban09}. This was based on  variations in EUV line widths of coronal plumes (viewed off
the solar limb) with height.
\begin{figure}
 \includegraphics[scale=0.3]{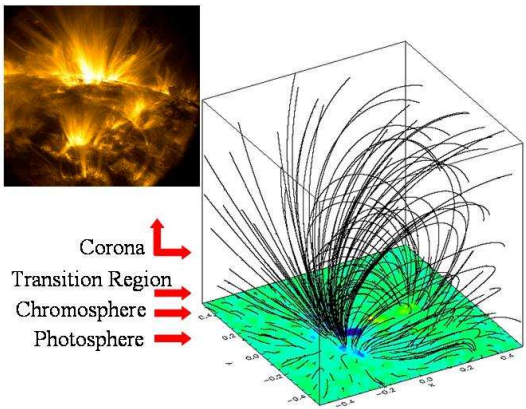}
  \caption{Top left panel shows TRACE 171 $\AA$ 
image of an active region in solar corona. The box shows potential
extrapolation (using Green's function method) of magnetic field lines based on an input from the measured normal component
of the photospheric magnetic field that can be used for numerical modelling.}
\end{figure}

Also, there seems to be is a serious gap in the
studies of the coronal heating problem, as here are no studies 
which consistently would monitor global
{\it (average) coronal temperature}.
There are studies that monitor TSI (total solar irradiance)
and its correlation to the (11 year-) solar activity cycle.
Recent study  by \citet{fro09} suggests that 
the long-term trend of TSI is most probably caused by a global temperature 
change of the Sun (photosphere) that does not influence the UV irradiance in 
the same way as the surface magnetic fields.
There is however, a work of \citet{asch01},
where they split the differential emission measure (DEM) into 10$^\circ$ sectors and estimate coronal
heating requirement sector-by-sector. For every full-disk dataset with EUV and
SXR coverage, it would be straightforward to compute a DEM of the entire
corona, from which an emission measure-weighted temperature could be extracted as
function of time. i.e. it could be possible to produce a graph of averaged (by all sectors i.e.
full disk) coronal  temperature as function of time, over
the time interval of e.g. $3 \times 11=33$ years with sufficient time cadence
(e.g. once per year or more). Such plot would answer a significant
question: whether the heat release in the corona is indeed related to the
magnetic field (which we know changes on 11 year timescale). There is certainly a good correlation
between the magnetic flux and SXR intensity (e.g. \citep{ben02}, or see Fig. 1.13 in 
\citet{asch2006} -- Indeed, there seems to be a good case for $I_{SXT}(t) \propto B^2$). However, it is
questionable whether $I_{SXT}(t)$ is a good proxy for global average coronal temperature, because
SXR emission is non thermal (produced by the forbidden lines) -- cf. Fig.2.3 from \citet{asch2006} showing
strong deviation of coronal emission from the blackbody spectrum.
The analogy with a patient who has caught cold is relevant here:
if a patient is sick, its temperature is routinely monitored by doctors. If we claim
there is a coronal heating problem, indeed,  
solar physicists should monitor the global average temperature 
of their patient -- the Sun's corona!

As far as modelling challenges are concerned, it seems a real progress can be
made if two-fluid simulation of heating release in AR is attempted.
When ion and electron dynamics is decoupled, we know that reconnection will
proceed fast \citep{kuz98,hesse99,birn01,pritchett01,th07,th08}
even without invoking anomalous resistivity.
What is encouraging is that we do not need to
use realistic ion to electron mass ration of $m_i/m_e=1836$.
As previous results have shown, as long as ion-electron dynamics is decoupled
simulation with $m_i/m_e=100$ is as good \citep{hesse01,th08}. This implies that significant reduction
in CPU requirements can be achieved.
The two-fluid simulation results then can be compared with the SDO's AIA and ROSA data to see whether inclusion of the 
two-fluid effects can provide fast enough reconnection or generated wave power 
and hence adequate coronal heating.

\section*{Acknowledgments}

Author is supported by the Science and Technology Facilities Council (STFC) of the United Kingdom.

\end{document}